%% file: main.tex
\documentclass{article}

\usepackage{arxiv}

\usepackage[utf8]{inputenc} 

\usepackage{amsfonts}       
\usepackage{nicefrac}       
\usepackage{microtype}      
\usepackage{lipsum}

\usepackage{color}
\usepackage{listings}
\usepackage{booktabs}
\usepackage{array}
\usepackage{graphicx}
\usepackage{longtable}
\usepackage{subfigure}

\usepackage{cite}
\usepackage{url}
\usepackage{fancyhdr}

\usepackage{soul}

\usepackage{float}
\usepackage{listings}
\usepackage{mdwmath}
\usepackage{mdwtab}
\usepackage{multirow}
\usepackage{multicol}
\usepackage{rotating}
\usepackage{setspace}
\usepackage[utf8]{inputenc}
\usepackage{lineno}
\usepackage{listings}

\usepackage{mdwmath}
\usepackage{mdwtab}
\usepackage{multirow}
\usepackage{multicol}
\usepackage{array}
\usepackage{booktabs}

\usepackage{enumitem}
\usepackage{xspace}
\usepackage[export]{adjustbox}
\usepackage{graphicx}
\usepackage{color,soul}
\usepackage{rotating}
\usepackage{setspace}
\usepackage{amsmath} 
\usepackage{amssymb}
\usepackage{float}
\usepackage{xcolor}

\usepackage{hyperref}
\usepackage{url}

\title{Blockchain-Enabled Accountability in Data Supply Chain: A Data Bill of Materials Approach}

\author{Yue Liu*, Dawen Zhang*, Boming Xia*, Julia Anticev†,
Tunde Adebayo†, Zhenchang Xing*, Moses Machao†\\
*Data61, CSIRO, Australia \\
†IM\&T, CSIRO, Australia
}

\begin{document}

\maketitle

\begin{abstract}
In the era of advanced artificial intelligence, highlighted by large-scale generative models like GPT-4, ensuring the traceability, verifiability, and reproducibility of datasets throughout their lifecycle is paramount for research institutions and technology companies. These organisations increasingly rely on vast corpora to train and fine-tune advanced AI models, resulting in intricate data supply chains that demand effective data governance mechanisms. In addition, the challenge intensifies as diverse stakeholders may use assorted tools, often without adequate measures to ensure the accountability of data and the reliability of outcomes. 
In this study, we adapt the concept of ``Software Bill of Materials" into the field of data governance and management to address the above challenges, and introduce ``Data Bill of Materials" (DataBOM) to capture the dependency relationship between different datasets and stakeholders by storing specific metadata. We demonstrate a platform architecture for providing blockchain-based DataBOM services, present the interaction protocol for stakeholders, and discuss the minimal requirements for DataBOM metadata. The proposed solution is evaluated in terms of feasibility and performance via case study and quantitative analysis respectively.

\end{abstract}

\textbf{Key terms - } Data governance, blockchain, accountability, data bill of materials, traceability, architecture

\section{Introduction}

Data governance is critical in the era of advanced artificial intelligence (AI), particularly with the proliferation of large-scale generative AI that necessitates extensive datasets for model training and fine-tuning. Organisations that navigate complex data supply chains involving multiple stakeholders and varied tools are facing challenges in ensuring the traceability, verifiability, and reproducibility of data. This complexity is compounded in cross-departmental or cross-organisational data exchanges, where maintaining data accountability becomes increasingly significant. This issue is exacerbated after the emergence of large-scale generative AI models such as Large Language Models (LLMs)~\cite{khan2022subjects}. As enterprises and research institutions all need large and high-quality corpora for model development and enhancement, the lack of effective governance frameworks to manage data creation, usage, and transfer, especially across diverse stakeholders, becomes evident.

Within a data supply chain, which involves continuing dataset artifact transformation and dissemination, stakeholders need to i) ensure data traceability in terms of the origin, authorisation and operations conducted on the dataset artifacts, ii) achieve data verifiability with authenticated sources and licence, iii) preserve data reproducibility that if questions are raised for specific steps on processing or transferring, and consequently, iv) the overall accountability to identify the responsible stakeholders if violations are detected. Nevertheless, current data governance models, often tied to specific platforms and focusing on data storage schemes (e.g., object storage, InterPlanetary File System), secure trading protocols ~\cite{8759960, 9324804}, and privacy regulations (e.g. the General Data Protection Regulation), fall short in addressing the dynamic nature of data flows from the perspective of the overall data supply chain and the requirement for platform-agnostic traceability solutions.

Hence, in this study, we adapt the concept of ``Software Bill of Materials" (SBOM) ~\cite{SBOM} into the field of data governance and management and explore ``Data Bill of Materials"~\cite{barclay2019towards} from the perspective of software architecture design. DataBOM provides a formal inventory of dependency relationships for datasets via specific metadata, facilitating traceability, verifiability, reproducibility, and ultimately accountability across the data supply chain. In addition, considering the decentralised nature of data supply chain, we employ blockchain technology to serve as an infrastructure for managing DataBOM among diverse stakeholders. The proposed solution is evaluated via a case study and performance analysis. The main contributions of this paper are as follows:

\begin{itemize}
    \item We introduce and leverage the concept of ``Data Bill of Materials" from the perspective of software architecture design, which provides a traceable, verifiable, and reproducible data supply chain to ensure accountability.
    
    \item We provide a three-tiered system architecture for Blockchain-based DataBOM, utilising on-chain smart contracts to create, retrieve and update DataBOM records, and discuss the application of different access control mechanisms and design patterns to enable fine-grained auxiliary services. To the best of our knowledge, this is the first study integrating Data Bill of Materials with blockchain technology in terms of software architecture design.

    \item We present the interaction protocol for operationalising the proposed architecture, and discuss the minimal metadata requirement for designing and deploying customised DataBOM smart contract.
\end{itemize}


\section{Background and Related Work}
\label{sec:background}

\subsection{Blockchain Technology}

Blockchain is essentially a distributed ledger technology which was popularised by the emergence of Bitcoin~\cite{Satoshi:bitcoin} and the subsequent cryptocurrencies. In recent years, blockchain has been leveraged as a software component for enabling decentralisation and on-chain autonomy in diverse applications including decentralised finance~\cite{CHEN2020e00151}, healthcare~\cite{MCGHIN201962}, etc., by providing two core elements: the underlying distributed ledger as immutable, transparent and secure data storage, and a decentralised ``computing" infrastructure facilitated through smart contracts. In this study, we explore blockchain-enabled Data Bill of Materials as the decentralised environment of data supply chain aligns with the nature of blockchain, while blockchain can provide on-chain program execution with a series of design patterns that can satisfy the assorted requirements of data security and privacy. Blockchain achieves data storage through encapsulating data within digital transactions, which serve as identifiable records that manage the versioning of data over time. The verification of digital transactions does not need to rely on any central authority to establish a trustworthy business relationship~\cite{scheuermann2015iacr}. Blockchain also affords Turing-complete on-chain programmability through smart contracts, which are user-defined programs deployed and executed on-chain. Smart contracts support advanced programming features such as triggers and conditions~\cite{Omohundro:2014} allowing the composition of complex business logic. 


\subsection{Bill of Materials}

The Bill of Materials (BOM), historically rooted in manufacturing, serves as a comprehensive list detailing all components necessary for product assembly~\cite{jiao2000generic}. This concept, pivotal for ensuring transparency and accountability in production, has been adapted to the realm of software development, manifesting as the Software Bill of Materials. An SBOM delineates all software components within an application, playing a crucial role in securing the software supply chain by cataloging every constituent component.
By extending BOM's principles to data, we aim to establish a DataBOM framework that ensures traceability and accountability throughout the data supply chain.

In the domain of SBOMs, three primary standard formats have emerged: Software Package Data Exchange (SPDX)\footnote{\url{https://spdx.dev/}}, CycloneDX\footnote{\url{https://cyclonedx.org/}}, and Software Identification (SWID) Tagging\footnote{\url{https://csrc.nist.gov/projects/Software-Identification-SWID}}.
SPDX, an open-source standard recognised internationally and hosted by the Linux Foundation, primarily focuses on licence compliance. CycloneDX, developed by OWASP in 2017, is tailored towards addressing security concerns. SWID Tagging, maintained by the US National Institute of Standards and Technology (NIST), aims to offer a robust mechanism for the transparent identification of software components.

Following the U.S. Executive Order on Improving the Nation’s Cybersecurity\footnote{\url{https://www.whitehouse.gov/briefing-room/presidential-actions/2021/05/12/executive-order-on-improving-the-nations-cybersecurity/}}, which mandates SBOMs for software procurement, there has been a marked acceleration in SBOM research and development. Studies have begun to explore SBOM's potential, challenges, and opportunities (e.g., \cite{Xia_Bi_Xing_Lu_Zhu_2023, stalnaker2024boms, bi2023way, nocera2023software}).
Notably, the notion of DataBOM, while nascent, was preliminarily explored in \cite{barclay2019towards}, which outlined a theoretical model for ensuring data traceability. The potential of DataBOM as a transformative tool for data governance has been further corroborated by subsequent studies, including \cite{stalnaker2024boms}, highlighting its capacity to address the challenges of data accountability and lifecycle management.


\subsection{Data Governance and Management}

The landscape of data governance and management has witnessed significant evolution, driven by both academic and industry efforts. 
Specifically for responsible AI data governance, frameworks such as datasheets \cite{gebru2021datasheets} and data cards \cite{pushkarna2022data} have been proposed for transparent data documentation via structured summaries of essential dataset facts from the perspective of a project lifecycle. On the other hand, Levin et al.~\cite{8818188} integrate AI with cloud object storage for maintaining the normal behaviour and health of IT Operations. 
Provena~\cite{Provena} leverages the W3C Provenance model for comprehensive workflow provenance in distributed environments.
Apache NiFi\footnote{\url{https://nifi.apache.org/}} 
underscores the importance of data provenance for operational analysis, facilitating the tracing of data through user-defined pipelines. This emphasis on provenance is further extended by initiatives like the Coalition for Content Provenance and Authenticity\footnote{\url{https://c2pa.org/}}, a collaborative effort by industry giants including Adobe and Microsoft, aimed at establishing standards for certifying the origin and history of media contents.

The exploration of decentralised technologies, such as the InterPlanetary File System and blockchain, introduces novel paradigms for data governance and management.
Research underscores the importance of a systematic approach to blockchain data governance, addressing key challenges in privacy, data quality assurance, and the provision of trustable data analytics~\cite{paik2019analysis}. 
Further enriching this discourse, the Secure Blockchain-Based Data Trading Ecosystem model~\cite{dai2019sdte} introduces a paradigm for secure data trading, emphasising governance mechanisms essential for safeguarding data integrity and confidentiality during transactions. Zhang et al.~\cite{zhang2023tag} propose a responsible web framework that leverages distributed ledger technology to empower users in managing consent and copyright of their online data.


In essence, the domain of data governance and management is transitioning towards a more interconnected, standardised, and user-centric approach, reflecting a collective move to address the pressing demands of data accountability. The concept of ``DataBOM" was introduced by Barclay et al.~\cite{barclay2019towards}, which is similar to data sheets and cards but focuses more on the changes and features of dataset artifact transformation and circulation. However, the extant work merely emphasised data traceability aspect, while this study extends the scope to include verifiability and reproducibility to further preserve accountability within data supply chain. 







\section{Blockchain-based Data Bill of Materials}
\label{sec:databom}

In this section, we present a platform architecture for blockchain-based DataBOM, the corresponding interaction protocol, and the minimal requirements for metadata contained in DataBOM records.

\begin{figure}[t]
	\centering
	\includegraphics[width=0.5\columnwidth]{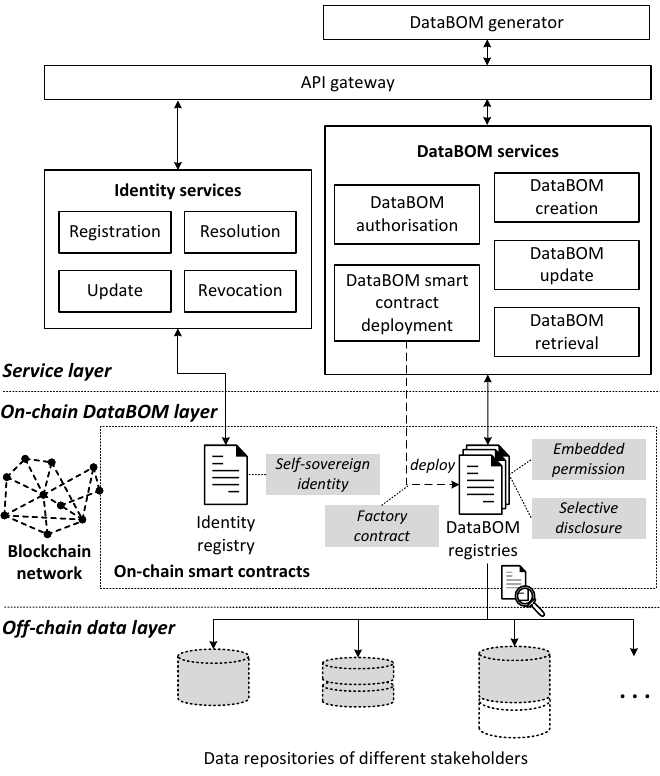}	
	\caption{Platform architecture for blockchain-based DataBOM.}
	\label{fig:architecture}
\end{figure}

\subsection{Architecture Design}

Fig.~\ref{fig:architecture} illustrates the overview of our proposed three-tiered architecture for blockchain-based DataBOM, including service layer, on-chain DataBOM layer, and off-chain data layer, and we also identify where a set of blockchain design patterns~\cite{9426788} can be applied. The \textit{service layer} consists of a \textit{DataBOM generator} to automatically detect and capture the metadata from updated dataset artifacts (e.g., Excel files), an \textit{API gateway}, \textit{identity services} and \textit{DataBOM services}. Specifically, \textit{identity services} manage stakeholders' on-chain identities for interaction, data source tracing, and accountability process, while \textit{DataBOM services} include the development and deployment of DataBOM registry smart contracts, the specific operations to DataBOM records (i.e., creation, update, retrieval), and the access authorisation to certain granularity levels of DataBOM.

The \textit{on-chain DataBOM layer} can process on-chain business logic via smart contracts. The \textit{identity registry} smart contract provides identity services, where \textit{self-sovereign identity} can be employed if the target usage scenario includes a series of cross-organisation interactions and collaborations, and hence the privacy requirements of both individuals and organisations should be noticed and satisfied by establishing formal business relationships. \textit{DataBOM registries} defines the data structure for selected dataset metadata, which can be utilised to identify, trace, verify and reproduce datasets along with the data supply chain. Considering the discrepancy of metadata and the different levels of granularity, customised \textit{DataBOM registries} are required to meet the diverse requirements for assorted datasets/projects. Hence, a \textit{factory contract} can serve as a template for generating customised contract instances. In a DataBOM registry, \textit{embedded permission} can grant access to stakeholders on the different granularity of a DataBOM record (i.e., only certain stakeholders can retrieve the records of a certain dataset artifact), whilst \textit{selective disclosure} enables further fine-grained visibility of confidential or sensitive metadata or certain time intervals. Finally, the \textit{off-chain data layer} consists of the data repositories, which are maintained by the stakeholders themselves instead of centralised in the implemented platform. The \textit{DataBOM registries} contains the paths directing to actual data repositories.

\begin{figure}[h!]
	\centering
	\includegraphics[width=0.5\columnwidth]{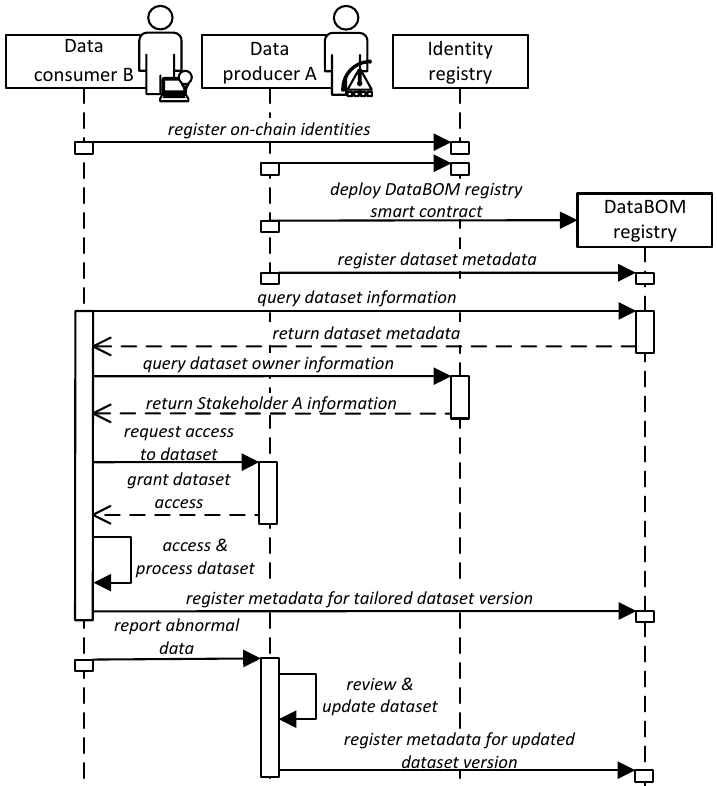}	
	\caption{Interaction protocol for blockchain-based DataBOM.}
	\label{fig:protocol}
\end{figure}

\subsection{Protocol for Blockchain-based DataBOM}


Fig.~\ref{fig:protocol} presents an interaction protocol for the stakeholders of blockchain-based DataBOM. First, all stakeholders need to register on-chain identities with their blockchain public keys. 
Data producer A possesses a dataset, thence deploys a \textit{DataBOM registry} smart contract for this dataset, and stores the selected metadata. 
Data consumer B is seeking suitable dataset(s) for certain projects, and queries the \textit{DataBOM registry} to understand the essential information of A's dataset. Afterwards, consumer B can retrieve producer A's identity information (e.g. service endpoint) from \textit{identity registry}, and request access to the dataset by directly contacting A. After authorisation, B accesses the dataset and decides to use only a subset of the included data. Hence they partition the dataset into multiple chunks, while the \textit{DataBOM generator} exports and registers metadata of this tailored dataset in \textit{DataBOM registry} as a new version. Please note that if the dataset identifier is changed within data supply chain, the dataset should be recorded in \textit{DataBOM registry} as a new dataset with the alternated identifier. If data consumer B discovers abnormal data in the dataset, they can report to the producer, who will review and update the dataset by addressing the issue, and then the updated dataset will trigger the \textit{DataBOM generator} to register the metadata in \textit{DataBOM registry} (e.g., the explanation for this new version).


\subsection{Minimal Requirements for DataBOM Metadata}

Since Data Bill of Materials is a new concept adopted from Software Bill of Materials, there is currently no widely accepted standard or guideline specifying what kind of dataset metadata should be included in DataBOM records. Hereby, we provide several insights on the minimal requirement for metadata considering the objectives of DataBOM. 

\textbf{Data traceability} refers to the ability to trace the flow of data throughout its lifecycle, from the origin or source to its final storage or use. It requires the identifiers of different datasets and the involved versions, and the dependency to other datasets or even particular versions to pinpoint the data source. \textbf{Data verifiability} denotes the ability to confirm the reliability and compliance of data. The generation, transfer and use of data should comply with relevant laws and specifications. Consequently, we recommend including the licence information in DataBOM to certify that the operations to a dataset or version do not violate the regulations. \textbf{Data reproducibility} stands for the ability to replicate the results of a certain phase within the data supply chain. In addition to the relationships with other datasets or versions, a stakeholder needs to specify the conducted operations to transform a dataset from the previous version to the current one. Other stakeholders can refer to this operation history for reproducing a precise dataset version. \textbf{Accountability} in data supply chain is embodied by the identifiability and answerability of stakeholders for their decisions and operations on dataset artifacts. The inherent public key infrastructure of blockchain and \textit{identity registry} enable the connection between stakeholders' identities and blockchain accounts. Hence, if violations are detected, it will be straightforward to locate the responsible stakeholder(s) according to dataset operation history and on-chain identities.

\section{Evaluation}
\label{sec:evaluation}

This section presents the proof-of-concept implementation and evaluation of the proposed blockchain-based DataBOM. We first illustrate how DataBOM can be integrated into the data supply chain of a research project consisting of four stakeholders, then elaborate the DataBOM data structure for this use case. In addition, we conduct experiments to evaluate the performance of creating and retrieving on-chain DataBOM records.

\begin{figure*}[t]
	\centering
	\includegraphics[width=\textwidth]{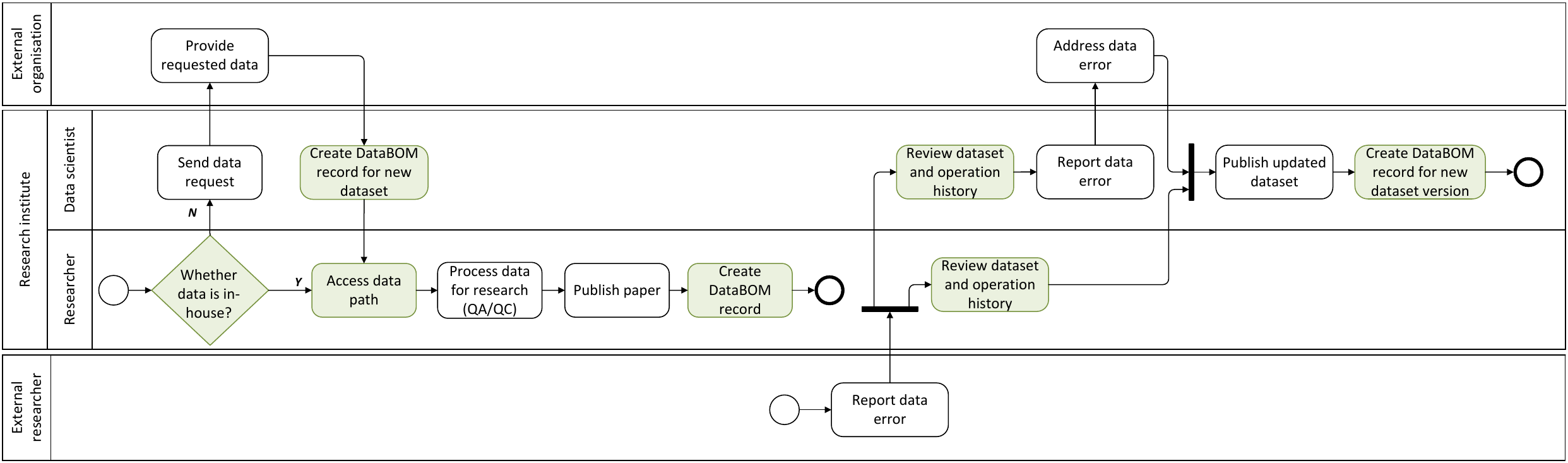}	
	\caption{Data supply chain in research projects.}
	\label{fig:case_study}
\end{figure*}

\subsection{Case Study}


Fig.~\ref{fig:case_study} illustrates a simplified data supply chain in a research project, where the annotated steps are supported by our proposed blockchain-based DataBOM services. Please note that we omit the DataBOM generator with automated metadata export and API invocation in the figure, which is implemented as a local file watcher in our minimal viable prototype. First, for a research project, a researcher can refer to \textit{DataBOM registries} as a catalogue to check whether the required dataset has been purchased by the institute. If the dataset is available, the researcher can directly access it via the data path recorded in DataBOM, otherwise, the researcher needs to contact the responsible data scientist, who will approach external organisations to acquire the dataset. After obtaining the target dataset, the data scientist stores it in data repository, and creates a DataBOM record for this new dataset for future searching and usage. Subsequently, the researcher uses the dataset for further analysis and may conduct specific operations, for instance, performing quality assurance/quality control, selecting a certain subset and rechunking this subset with other datasets, etc. After the research paper publication, the researcher can create a DataBOM record for the revised dataset. In particular, the paper's Digital Object Identifier (DOI) can be added as a data path considering the paper itself provides the link to data repository, and explains the methodology for processing the dataset. Other researchers can review the published paper and dataset (in this case the tailored dataset is open-source). If there are some errors found in the dataset (e.g., missing data should be set to a specific value but was found to be 0), the external researcher can report the error to the corresponding research institute. The researcher and data scientist both need to review the dataset and operation history recorded in DataBOM, while the data scientist also needs to notify the upstream organisation about the found errors. Regarding addressing the errors, the data scientist can publish the updated dataset and create a DataBOM record explaining this new version, which can be retrieved and used in future projects within the research institute.


We explored the data repository provided by the data scientist in our case study, and investigated the dataset metadata. We extracted and selected specific global attributes in the dataset artifacts to be included in the \textit{DataBOM registry} smart contract, including Universally Unique Identifier (UUID), file name, licence, summary, and history, etc. The DataBOM data structure contains four structs to simulate the real-world file system and enable different granularity, including project, year, dataset, and version. An individual research project covers multiple years, while each year contains a series of dataset artifacts, and each dataset artifact may have multiple versions. Specifically, the authorisation lists are enabled by \textit{embedded permission} for project and dataset levels respectively, in such manner, only authorised stakeholders at the project level can create new dataset records, and the authorised ones at the dataset level can create new versions for a particular dataset. Considering dataset artifacts are identified via UUID, if multiple dataset artifacts are merged and a new UUID is used as the identifier, the resulted artifact should be recorded as a new dataset.

\subsection{Performance Analysis}

We conducted quantitative experiments to measure the throughput and response time of two critical DataBOM service APIs in the proposed architecture: \textit{DataBOM creation} and \textit{DataBOM retrieval}, while other services have similar performance as they are all essentially the operations to store and query data to/from on-chain smart contracts. We implemented a minimal viable prototype for proof of concept of our proposed solution. The prototype API gateway is developed using Node.js v18.19.0, Web3.js v1.10.4 and Solidity v0.5.3, and it can support input and output in JSON format. The on-chain DataBOM layer is deployed on a local Ganache 
blockchain network. We configured the Ganache network block size to simulate the Ethereum Mainnet. The API requests, containing the corresponding required information of the four services, are produced via JMeter. JMeter is configured to 20 creations per batch (API calls). Each test ran for 30 minutes. Secondly, we tested the APIs 1000 times, and recorded the response time.

\begin{figure}[t]
	\centering
	\includegraphics[width=0.5\columnwidth]{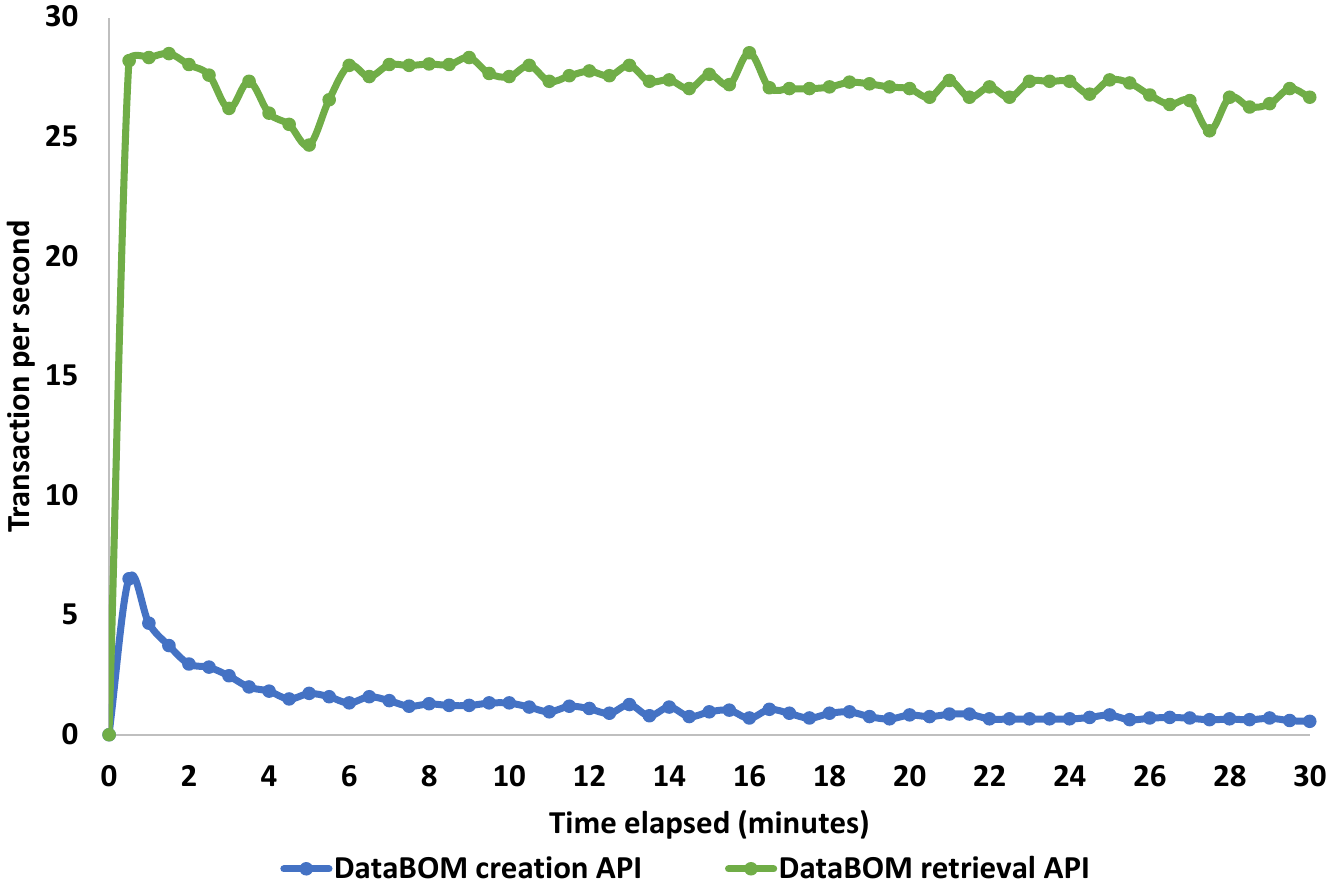}	
	\caption{Throughput for two RESTful APIs.}
	\label{fig:tps}
\end{figure}

\begin{figure}[t]
	\centering
	\includegraphics[width=0.5\columnwidth]{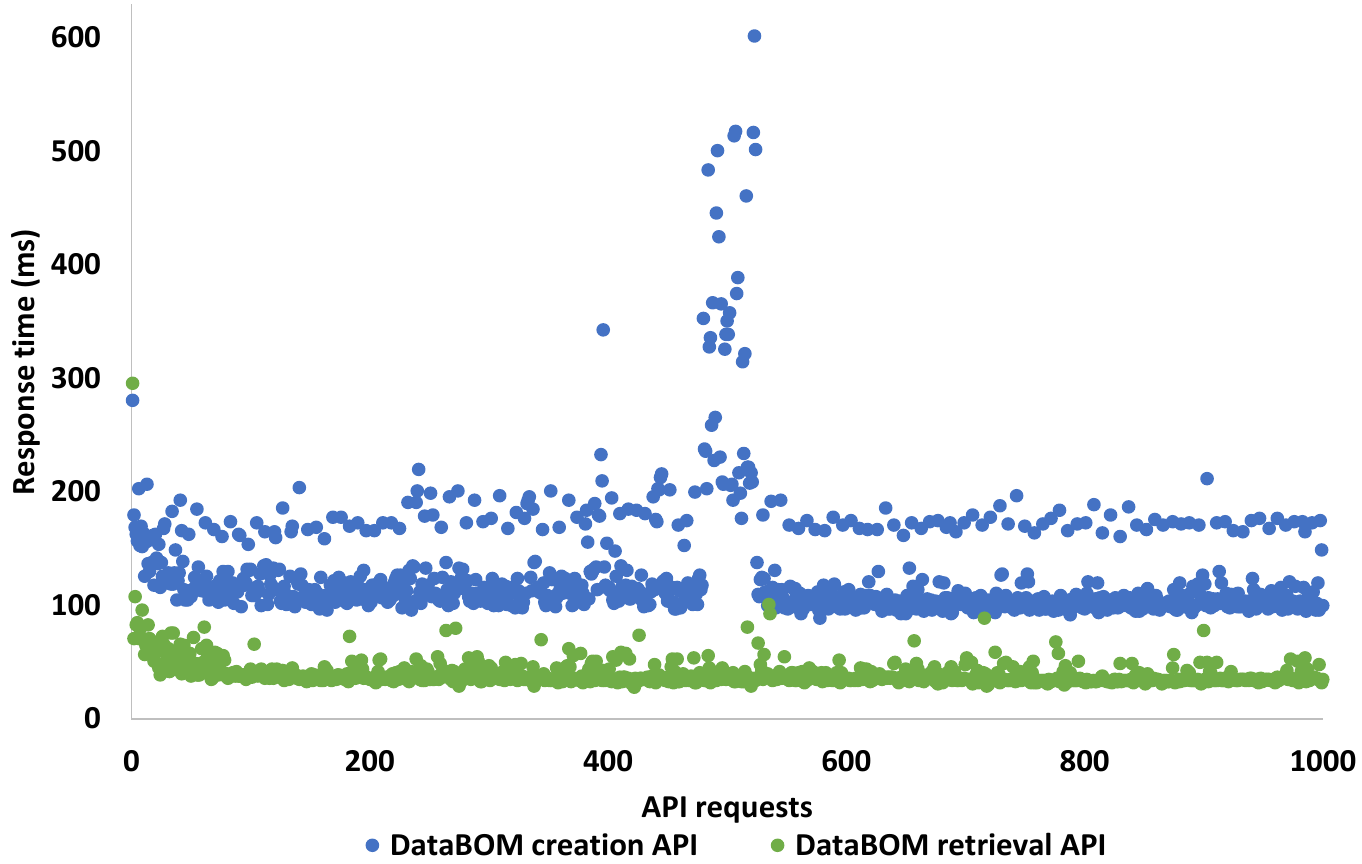}	
	\caption{Response time for two RESTful APIs.}
	\label{fig:response}
\end{figure}

Fig.~\ref{fig:tps} illustrates the measurement results for \textit{DataBOM creation} and \textit{DataBOM retrieval} API throughput. The x-axis represents the time elapsed since the start of the experiment (in minutes), while the y-axis represents the average throughput (transaction per second - tps). It can be observed that \textit{DataBOM retrieval} throughput stayed around 27 tps, while \textit{DataBOM creation} throughput peaked at the beginning (6.5 tps), but decreased to 1 tps afterwards. The performance of \textit{DataBOM creation} is heavily dependent on the transaction and block size, since the DataBOM information can be large by including the licence, summary, and operation history, whilst the block size is fixed. 
Fig.~\ref{fig:response} demonstrates the response time of running the two APIs for 1000 times respectively. The x-axis represents the number of API requests (1 to 1000), and the y-axis represents the response time of each sent request (in milliseconds - ms). For \textit{DataBOM creation} API, the majority of requests are completed between 100~200 ms, while a small subset requires longer response time, whilst most \textit{DataBOM retrieval} API requests are all carried out within 100 ms except the first several ones. We assume this is caused by the communication overhead between Jmeter, API gateway and Ganache blockchain at the beginning of the test.

\section{Conclusion}
\label{sec:conclusion}

This paper presents a platform architecture of blockchain-based data bill of materials. In the design, the identity services enable users to manage their on-chain identities, and the DataBOM services facilitate data traceability, verifiability, reproducibility and overall accountability by storing the metadata of dataset artifacts in on-chain smart contracts. 
We introduce the interaction protocol of our proposed architecture, and discuss the minimal requirements for DataBOM metadata. A minimal viable prototype for proof-of-concept is implemented, and we perform evaluation regarding feasibility and performance. In the future, we plan to: i) explore the feasibility of utilising foundation model based agents in the DataBOM generator to automatically analyse the dataset artifacts and extract required metadata, and; ii) integrate SBOM and DataBOM in the field of AI governance, to operationalise AIBOM~\cite{xia2023trust}.

\input{main.bbl}



\end{document}

%% file: main.bbl

%% file: main.bbl
\begin{thebibliography}{10}
\providecommand{\url}[1]{#1}
\csname url@samestyle\endcsname
\providecommand{\newblock}{\relax}
\providecommand{\bibinfo}[2]{#2}
\providecommand{\BIBentrySTDinterwordspacing}{\spaceskip=0pt\relax}
\providecommand{\BIBentryALTinterwordstretchfactor}{4}
\providecommand{\BIBentryALTinterwordspacing}{\spaceskip=\fontdimen2\font plus
\BIBentryALTinterwordstretchfactor\fontdimen3\font minus \fontdimen4\font\relax}
\providecommand{\BIBforeignlanguage}[2]{{%
\expandafter\ifx\csname l@#1\endcsname\relax
\typeout{** WARNING: IEEEtran.bst: No hyphenation pattern has been}%
\typeout{** loaded for the language `#1'. Using the pattern for}%
\typeout{** the default language instead.}%
\else
\language=\csname l@#1\endcsname
\fi
#2}}
\providecommand{\BIBdecl}{\relax}
\BIBdecl

\bibitem{khan2022subjects}
M.~Khan and A.~Hanna, ``The subjects and stages of ai dataset development: A framework for dataset accountability,'' Available at SSRN: https://ssrn.com/abstract=4217148 or http://dx.doi.org/10.2139/ssrn.4217148, 2022, accessed 17-January-2024.

\bibitem{8759960}
W.~Dai, C.~Dai, K.-K.~R. Choo, C.~Cui, D.~Zou, and H.~Jin, ``Sdte: A secure blockchain-based data trading ecosystem,'' \emph{IEEE Transactions on Information Forensics and Security}, vol.~15, pp. 725--737, 2020.

\bibitem{9324804}
L.~D. Nguyen, I.~Leyva-Mayorga, A.~N. Lewis, and P.~Popovski, ``Modeling and analysis of data trading on blockchain-based market in iot networks,'' \emph{IEEE Internet of Things Journal}, vol.~8, no.~8, pp. 6487--6497, 2021.

\bibitem{SBOM}
T.~U. S.~D. of~Commerce, ``The minimum elements for a software bill of materials,'' \url{https://www.ntia.doc.gov/files/ntia/publications/sbom_minimum_elements_report.pdf}, accessed 17-January-2024.

\bibitem{barclay2019towards}
I.~Barclay, A.~Preece, I.~Taylor, and D.~Verma, ``Towards traceability in data ecosystems using a bill of materials model,'' \emph{arXiv preprint arXiv:1904.04253}, 2019.

\bibitem{Satoshi:bitcoin}
S.~Nakamoto, ``Bitcoin: A peer-to-peer electronic cash system,'' \url{https://bitcoin.org/bitcoin.pdf}, 2008, accessed 6-June-2022.

\bibitem{CHEN2020e00151}
\BIBentryALTinterwordspacing
Y.~Chen and C.~Bellavitis, ``Blockchain disruption and decentralized finance: The rise of decentralized business models,'' \emph{Journal of Business Venturing Insights}, vol.~13, p. e00151, 2020. [Online]. Available: \url{https://www.sciencedirect.com/science/article/pii/S2352673419300824}
\BIBentrySTDinterwordspacing

\bibitem{MCGHIN201962}
\BIBentryALTinterwordspacing
T.~McGhin, K.-K.~R. Choo, C.~Z. Liu, and D.~He, ``Blockchain in healthcare applications: Research challenges and opportunities,'' \emph{Journal of Network and Computer Applications}, vol. 135, pp. 62--75, 2019. [Online]. Available: \url{https://www.sciencedirect.com/science/article/pii/S1084804519300864}
\BIBentrySTDinterwordspacing

\bibitem{scheuermann2015iacr}
F.~Tschorsch and B.~Scheuermann, ``Bitcoin and beyond: {A} technical survey on decentralized digital currencies,'' \emph{IEEE Communications Surveys \& Tutorials}, vol.~18, no.~3, p. 464, 2016.

\bibitem{Omohundro:2014}
S.~Omohundro, ``Cryptocurrencies, smart contracts, and artificial intelligence,'' \emph{AI Matters}, vol.~1, no.~2, pp. 19--21, Dec. 2014.

\bibitem{jiao2000generic}
J.~Jiao, M.~M. Tseng, Q.~Ma, and Y.~Zou, ``Generic bill-of-materials-and-operations for high-variety production management,'' \emph{Concurrent Engineering}, vol.~8, no.~4, pp. 297--321, 2000.

\bibitem{Xia_Bi_Xing_Lu_Zhu_2023}
B.~Xia, T.~Bi, Z.~Xing, Q.~Lu, and L.~Zhu, ``An empirical study on software bill of materials: Where we stand and the road ahead,'' in \emph{2023 IEEE/ACM 45th International Conference on Software Engineering (ICSE)}, 2023, pp. 2630--2642.

\bibitem{stalnaker2024boms}
T.~Stalnaker, N.~Wintersgill, O.~Chaparro, M.~Di~Penta, D.~M. German, and D.~Poshyvanyk, ``Boms away! inside the minds of stakeholders: A comprehensive study of bills of materials for software systems,'' in \emph{Proceedings of the 46th IEEE/ACM International Conference on Software Engineering}, 2024, pp. 1--13.

\bibitem{bi2023way}
T.~Bi, B.~Xia, Z.~Xing, Q.~Lu, and L.~Zhu, ``On the way to sboms: Investigating design issues and solutions in practice,'' \emph{arXiv preprint arXiv:2304.13261}, 2023.

\bibitem{nocera2023software}
S.~Nocera, S.~Romano, M.~Di~Penta, R.~Francese, and G.~Scanniello, ``Software bill of materials adoption: A mining study from github,'' in \emph{2023 IEEE International Conference on Software Maintenance and Evolution (ICSME)}.\hskip 1em plus 0.5em minus 0.4em\relax IEEE, 2023, pp. 39--49.

\bibitem{gebru2021datasheets}
T.~Gebru, J.~Morgenstern, B.~Vecchione, J.~W. Vaughan, H.~Wallach, H.~D. Iii, and K.~Crawford, ``Datasheets for datasets,'' \emph{Communications of the ACM}, vol.~64, no.~12, pp. 86--92, 2021.

\bibitem{pushkarna2022data}
M.~Pushkarna, A.~Zaldivar, and O.~Kjartansson, ``Data cards: Purposeful and transparent dataset documentation for responsible ai,'' in \emph{Proceedings of the 2022 ACM Conference on Fairness, Accountability, and Transparency}, 2022, pp. 1776--1826.

\bibitem{8818188}
A.~Levin, S.~Garion, E.~K. Kolodner, D.~H. Lorenz, K.~Barabash, M.~Kugler, and N.~McShane, ``Aiops for a cloud object storage service,'' in \emph{2019 IEEE International Congress on Big Data (BigDataCongress)}, 2019, pp. 165--169.

\bibitem{Provena}
J.~Yu, P.~Baker, S.~J. Cox, R.~Petridis, A.~C. Freebairn, F.~Mirza, L.~Thomas, S.~Tickell, D.~Lemon, and M.~Rezvani, ``Provena: A provenance system for large distributed modelling and simulation workflows,'' in \emph{25th International Congress on Modelling and Simulation (MODSIM2023)}, 2023, pp. 14--20.

\bibitem{paik2019analysis}
H.-Y. Paik, X.~Xu, H.~D. Bandara, S.~U. Lee, and S.~K. Lo, ``Analysis of data management in blockchain-based systems: From architecture to governance,'' \emph{Ieee Access}, vol.~7, pp. 186\,091--186\,107, 2019.

\bibitem{dai2019sdte}
W.~Dai, C.~Dai, K.-K.~R. Choo, C.~Cui, D.~Zou, and H.~Jin, ``Sdte: A secure blockchain-based data trading ecosystem,'' \emph{IEEE Transactions on Information Forensics and Security}, vol.~15, pp. 725--737, 2019.

\bibitem{zhang2023tag}
D.~Zhang, B.~Xia, Y.~Liu, X.~Xu, T.~Hoang, Z.~Xing, M.~Staples, Q.~Lu, and L.~Zhu, ``Tag your fish in the broken net: A responsible web framework for protecting online privacy and copyright,'' \emph{arXiv preprint arXiv:2310.07915}, 2023.

\bibitem{9426788}
X.~Xu, H.~Dilum~Bandara, Q.~Lu, I.~Weber, L.~Bass, and L.~Zhu, ``A decision model for choosing patterns in blockchain-based applications,'' in \emph{2021 IEEE 18th International Conference on Software Architecture (ICSA)}, 2021, pp. 47--57.

\bibitem{xia2023trust}
B.~Xia, D.~Zhang, Y.~Liu, Q.~Lu, Z.~Xing, and L.~Zhu, ``Trust in software supply chains: Blockchain-enabled sbom and the aibom future,'' \emph{arXiv preprint arXiv:2307.02088}, 2023.

\end{thebibliography}
